\documentclass[12pt]{article}
\usepackage[margin=2cm]{geometry}
\usepackage{setspace}
\usepackage[numbers,sort&compress]{natbib}
\usepackage{xcolor}
\usepackage{graphicx}
\usepackage{dcolumn}
\usepackage{bm}
\usepackage{amsmath,amssymb}
\usepackage{hyperref}
\usepackage{xr}
\makeatletter
\newcommand*{\addFileDependency}[1]{
  \typeout{(#1)}
  \@addtofilelist{#1}
  \IfFileExists{#1}{}{\typeout{No file #1.}}
}
\makeatother
 
\newcommand*{\myexternaldocument}[1]{%
    \externaldocument{#1}%
    \addFileDependency{#1.tex}%
    \addFileDependency{#1.aux}%
}
\myexternaldocument{SI_nanoscale_revised_II}

\usepackage{mathtools}

\newcommand{\XYZ }{ }

\definecolor{myblue}{RGB}{17, 30, 108}
\definecolor{myred}{RGB}{184, 15, 10}

\title{The photothermal nonlinearity in plasmon-assisted photocatalysis}

\begin{document}

\author{Ieng Wai Un,$^{a\ast}$, Yonatan Dubi$^{b}$ and Yonatan Sivan$^{a}$
\\
\normalsize{$^{a}$School of Electrical and Computer Engineering, Ben-Gurion University of the Negev, Israel}\\
\normalsize{$^{b}$Department of Chemistry, Ben-Gurion University of the Negev, Beer-Sheva, 8410501, Israel.}\\
\normalsize{$^\ast$E-mail: iengwai@post.bgu.ac.il}}

\maketitle

\newpage
\begin{abstract}
Understanding the intricate relation between illumination and temperature in metallic nano-particles is crucial for elucidating the role of illumination in various physical processes which rely on plasmonic enhancement but are also sensitive to temperature.  Recent studies have shown that the temperature rise in optically-thick ensembles of metal nanoparticles under intense illumination is dominated by the thermal conductivity of the host, rather than by the optical properties of the metal or the host. Here, we show that the temperature dependence of the thermal conductivity of the host dominates the nonlinear photothermal response of these systems. In particular, this dependence typically causes the temperature rise to become strongly sublinear, reaching even several tens of percent. We then show that this can explain experimental observations in several recent plasmon-assisted photocatalysis experiments. This shows that any claim for dominance of non-thermal electrons in plasmon-assisted photocatalysis must account first for this photothermal nonlinear mechanism.
\end{abstract}

\section{Introduction}
The use of illuminated metallic surfaces to enhance the yield of chemical reactions (aka plasmon-assisted photocatalysis) was proposed by Nitzan and Brus already in 1981~\cite{Nitzan_PAPC_1981} and implemented experimentally shortly later~\cite{Osgood_PAPC_1983}. However, only after several decades of slow progress (see e.g., Refs.~\citenum{Mirkin_PAPC_2001,Brus_PAPC_2003,Tatsuma_2005,plasmonic_photo_synthesis_Misawa_2008,Kitamura_PAPC_2009,Quidant_PAPC_2012}), this line of research has rapidly gained popularity following several high impact publications (see, e.g. Refs.~\citenum{plasmonic_photocatalysis_Clavero,plasmonic-chemistry-Baffou,Valentine_hot_e_review,plasmonic_photocatalysis_Chen-Wang} for some recent reviews). The growing interest was propelled by claims in some of the more famous papers on the topic~\cite{plasmonic_photocatalysis_1,Halas_dissociation_H2_TiO2,plasmonic_photocatalysis_Linic,Halas_H2_dissociation_SiO2,Halas_Science_2018} that the reaction rate increases due to the excitation of high energy non-thermal electrons in the metal (aka ``hot'' electrons), which then tunnel out of the metal, and provide the necessary energy for the reactants to allow them to be converted into the products more efficiently. 

However, these very papers (Refs.~\citenum{plasmonic_photocatalysis_1,Halas_dissociation_H2_TiO2,plasmonic_photocatalysis_Linic,Halas_H2_dissociation_SiO2,Halas_Science_2018}) were shown 
to suffer from technical and conceptual flaws (including improper temperature measurements, improper data normalization etc., see discussion in Refs.~\citenum{anti-Halas-Science-paper,R2R,Y2-eppur-si-riscalda,Baffou-Quidant-Baldi,anti-Halas-NatCat-paper}). Instead, a purely thermal mechanism was shown to be able to explain the experimental data quite convincingly~\cite{anti-Halas-Science-paper,Y2-eppur-si-riscalda,Dubi-Sivan-APL-Perspective,anti-Halas-NatCat-paper}. In particular, a shifted Arrhenius Law for the reaction rate, $R \sim \exp\left(-\frac{ \mathcal{E}_a}{k_B T({\bf r}) + a I_{\textrm{inc}}}\right)$ whereby the temperature of the system was corrected for the illumination-induced heating was shown to provide an excellent fit to the published data, essentially with no fit parameters. This result was corroborated with the first ever complete calculation of the steady-state electron non-equilibrium in metals~\cite{Dubi-Sivan}, a consequent Fermi golden-rule argument~\cite{Dubi-Sivan-Faraday} that pointed to the improbability of nonthermal electrons to cause the catalysis, and by detailed thermal simulations where the dynamics of the heat generated from each of the nanoparticles (NPs) in the system was properly modelled~\cite{Y2-eppur-si-riscalda}. Similar criticism was raised in an independent study~\cite{Baffou-Quidant-Baldi}, and characterization which is inline with our approach was employed by several groups (see, e.g., Refs.~\citenum{Baldi-ACS-Nano-2018,Boltasseva_LPR_2020,yu2019plasmonic,Yugang_Sun,Baldi-Nat-Comm-2020}).

In a consequent paper~\cite{Un-Sivan-sensitivity}, it was shown that in many typical configurations, the tedious modelling of the contributions of each of the heated NPs in the sample can be replaced by an effective medium approximation. This approach also enabled accounting for the exact reactor geometry, constituent materials and boundary conditions, thus, enabling a quantitative comparison to the measured data (see Fig.~\ref{fig:scheme}(a)). This series of works was lately extended to account also for fluid dynamics effects and for redox reactions~\cite{IW-Sivan-redox-paper} (Fig.~\ref{fig:scheme}(b)). 

The bottom line of the thermal modelling was that when attempting to quantitatively separate thermal and non-thermal effects in plasmon-assisted photocatalysis experiments, one has to overcome a conceptual difficulty - the thermocatalysis control experiments must reproduce the {\em exact spatially non-uniform temperature profile} induced by the illumination, otherwise, when subtracting the thermocatalysis rate from the photocatalysis rate (e.g., as in Refs.~\citenum{Halas_Science_2018,Liu-Everitt-Nano-research-2019}), any difference between the temperature distributions in an inaccurate control and the corresponding photocatalysis experiment is bound to be incorrectly interpreted as ``hot'' electron action. This issue is particularly important because the Arrhenius Law shows that the reaction rate has an exponential sensitivity to the temperature distribution~\cite{Y2-eppur-si-riscalda}. 

Detailed measurements and/or calculations of the temperature distribution in the studied samples indeed constituted a central role in several recent demonstrations of non-thermal effects in plasmon-assisted photocatalysis~\cite{Baldi-ACS-Nano-2018,Liu-Everitt-Nano-research-2019,Cortes_Nano_lett_2019,Boltasseva_LPR_2020,yu2019plasmonic}. However, while the simple thermal calculations done so far were sufficient for relatively simple scenarios, they may not be sufficient to account for more complicated ones. Those include, in particular, high intensity illumination which invokes {\XYZ steady-state} nonlinear thermo-optic and photothermal effects~\cite{Donner_thermal_lensing,plasmonic-SAX-PRL,Sivan-Chu-high-T-nl-plasmonics,Gurwich-Sivan-CW-nlty-metal_NP,IWU-Sivan-CW-nlty-metal_NP}, which are usually simply ignored (without any justification). Already in the context of thin metal layers~\cite{Wilson_deps_dT,Shalaev_ellipsometry_silver,Shalaev_ellipsometry_gold,PT_Shen_ellipsometry_gold} and single NPs~\cite{plasmonic-SAX-PRL,Hashimoto-plasmon-heat-fab-glass-2016,Stoll_review,Sivan-Chu-high-T-nl-plasmonics,Gurwich-Sivan-CW-nlty-metal_NP,IWU-Sivan-CW-nlty-metal_NP}, this effect was shown to cause deviations of several tens to hundreds of percent in the permittivity, and hence in the field and temperature distributions compared with the purely uniform linear thermal response. Note that such a nonlinear effect is far greater than conventional nonlinear optical effects.

In this work, we go beyond the study of the linear response, and evaluate the importance of nonlinear photothermal effects in large random ensembles of metal NPs, suitable to plasmon-assisted photocatalysis experiments {\XYZ (Fig.~\ref{fig:scheme}(a)-(b)}, but also to many other types of experiments in nonlinear optics (see e.g., Refs.~\citenum{Smith_Boyd_EMT_kerr_media,Boyd-metal-nlty,Palpant_EMT_kerr_media,Khurgin_chi3,Donner_thermal_lensing,Blau_review_SA_RSA}). First, in Section~\ref{sec:qualitative} we provide a qualitative analysis that points to the {\XYZ most important parameter that affects the overall nonlinear photothermal response of typical plasmon-assisted photocatalysis systems, namely, the thermal conductivity of the host. We also identify those parameters which have a negligible effect on the nonlinearity, namely, the optical parameters. Then, in Section~\ref{sec:methodology} we switch to a rigorous analysis and describe} the methodology we employ to calculate the temperature distribution in the samples considered. The qualitative analysis is then {\XYZ applied in Section~\ref{sec:experiments} to two sets of experimental data taken from recent high-impact plasmon-assisted photocatalysis experiments. The good match of our analysis with the experimental data} indicates that the photothermal nonlinearity is indeed a significant effect, such that a neglect to account for it is bound to lead to an overestimate of the role of non-thermal electrons. Section~\ref{sec:discussion} provides a discussion and outlook.

\section{A qualitative analysis}\label{sec:qualitative}
In order to achieve a qualitative understanding of the high temperature and/or intensity response of plasmon-assisted photocatalysis systems (i.e., the photothermal nonlinearity), we start by considering a simplified configuration, namely, we assume that the sample consists of metal NPs (with dielectric permittivity $\varepsilon_m = \varepsilon_m^{\prime} + i\varepsilon_m^{\prime\prime}$, thermal conductivity $\kappa_m$ and NP number density $n_p$) distributed in a cylinder-shape volume and immersed in a uniform host material (with thermal conductivity $\kappa_h$). 

In our previous work~\cite{Un-Sivan-sensitivity}, we have shown that in the weak illumination limit, the temperature rise at the top center of such a sample can be approximately written as
\begin{align}\label{eq:Delta_T_top_approx}
\Delta T^\textrm{top} \approx \dfrac{I_\textrm{inc}\rho_b}{2\kappa_h}\left(1 - e^{-H/\delta_\textrm{skin}} \right),
\end{align}
where $\rho_b$ is the beam radius, $H$ is the sample thickness, and $\delta_\textrm{skin}$ is the penetration (skin) depth of light in the catalyst sample. The inverse of the penetration (skin) depth (i.e., the absorption coefficient) is related to the NP number density $n_p$ and the absorption cross-section $\sigma_\textrm{abs}$ via
\begin{align}\label{eq:skin_depth}
1/\delta_\textrm{skin}(\omega) = n_p \sigma_\textrm{abs}(\omega).
\end{align}

\begin{figure*}[h]
\centering
\includegraphics[width=1\textwidth]{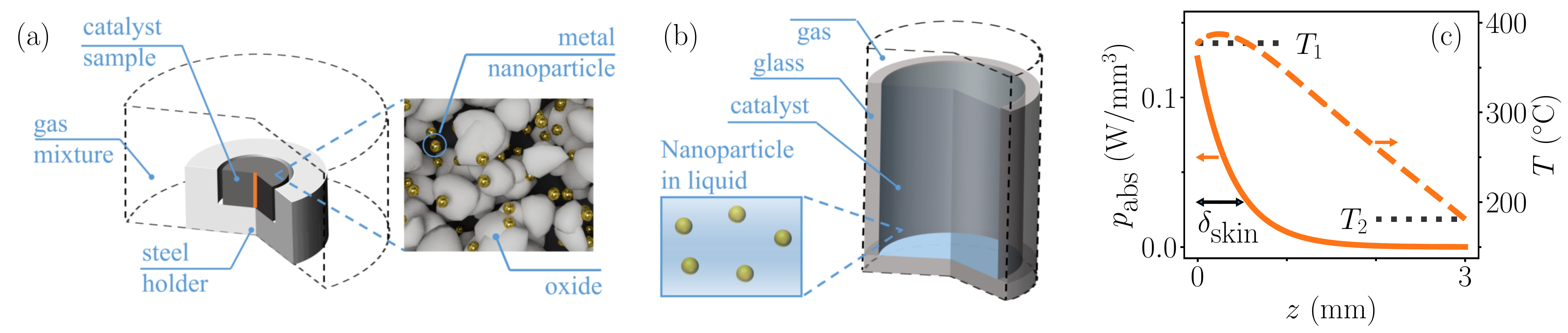}
\caption{(Color online) {\XYZ (a) A schematic illustration of a photocatalytic chamber (left) and of a pellet-based catalyst sample (right). Note the sparsity of the metal nanoparticles and the near-percolation nature of the micron-size oxide particles. (b) A schematic illustration of a typical plasmon-assisted photocatalysis setup based on metal nanoparticles liquid suspension. In both set-ups, a heat flux boundary condition is applied to all domain boundaries (marked by dashed lines). (c)} The heat source density (orange solid line) and temperature (orange dashed line) along the optical axis into the catalyst sample (labeled by the orange solid line in (a)). The black dotted lines represent the top ($T_1$) and the bottom ($T_2$) surface temperature.} \label{fig:scheme}
\end{figure*}

{\XYZ We now recall that in plasmon-assisted photocatalysis experiments, the light penetration depth is usually designed to be smaller than the sample thickness (i.e., such that $H \ll \delta_{\textrm{skin}}(\omega)$ for all wavelengths in the illumination) to ensure that all the illumination energy is absorbed, see Fig.~\ref{fig:scheme}(c). 
Eq.~(\ref{eq:Delta_T_top_approx}) shows that under such conditions}, the temperature rise simply becomes $\Delta T^{\textrm{top}} \approx \dfrac{I_{\textrm{inc}}\rho_b}{2\kappa_h}$, so that the overall temperature rise is weakly-sensitive to the illumination spectrum, NP shape, size, and density, but exhibits an inverse proportion to the host thermal conductivity~\cite{Un-Sivan-sensitivity}. 

When the illumination intensity is increased, the illumination-induced heating of the NPs causes a modification of the optical and thermal properties of the NPs and their surrounding. This effect gives rise to a nonlinear dependence of the sample temperature on the illumination intensity. {\XYZ Eq.~\eqref{eq:Delta_T_top_approx} and its approximation shows that the photothermal nonlinearity would be dominated by the temperature dependence of the thermal conductivity of the host, $\kappa_h$. To understand this potentially non-intuitive result, we note that} the increase of the imaginary part of the metal permittivity ($\varepsilon_m^{\prime\prime}$) with temperature reduces the quality factor of the plasmonic resonance of the NPs~\cite{Sivan-Chu-high-T-nl-plasmonics,Gurwich-Sivan-CW-nlty-metal_NP,IWU-Sivan-CW-nlty-metal_NP}. Meanwhile, the change of the real part of the metal permittivity ($\varepsilon_m^{\prime}$) causes a resonance shift of the absorption spectrum. Although the impact of the change of the metal permittivity on the absorption cross-section varies in a complex manner with the NP size and the illumination wavelength, see e.g. Refs.~\citenum{Sivan-Chu-high-T-nl-plasmonics,Gurwich-Sivan-CW-nlty-metal_NP, IWU-Sivan-CW-nlty-metal_NP}, and although the sensitivity of the metal permittivity to the rising temperature is relatively high~\cite{Rosei_Au_diel_function2,Winsemius_Au_Ag_Cu_ellipsometry,Palpant_EMT_kerr_media,Shalaev_ellipsometry_silver,Shalaev_ellipsometry_gold,Sivan-Chu-high-T-nl-plasmonics,PT_Shen_ellipsometry_gold}, the metal permittivity has a relatively small contribution to the overall photothermal response of the sample when the penetration depth is much thinner than the sample thickness, as shown in Eq.~\eqref{eq:Delta_T_top_approx} {\XYZ because the sample absorbs all light regardless of these changes}. Moreover, since the temperature dependence of the optical properties of the host has a similar effect on the absorption cross-section as the real part of the metal permittivity~\cite{Gurwich-Sivan-CW-nlty-metal_NP,IWU-Sivan-CW-nlty-metal_NP} (i.e., it causes a resonance shift), the nonlinear response due to the change of the host permittivity is also small. 

On the other hand, the contribution of the thermal properties to the overall photothermal nonlinearity is significant; it naturally depends on the volume fraction of the various materials. Because the metal occupies a small fraction of the sample volume, one can appreciate that the change of the metal thermal properties hardly contributes to the nonlinear response of the sample. In contrast, Eq.~(\ref{eq:Delta_T_top_approx}) shows that the thermal properties of the {\em host} matters much more {\XYZ (be it gas, liquid, a porous composite or even a (dielectric) solid)}. Clearly, an illumination-induced increase of the thermal conductivity with the temperature means that the overall temperature rise in the sample would become sublinear as a function of the illumination intensity. Judging by the typical thermoderivative of these properties, the nonlinearity is expected to manifest itself at a temperature rise of several hundreds of degrees, see Refs.~\citenum{Gurwich-Sivan-CW-nlty-metal_NP,IWU-Sivan-CW-nlty-metal_NP}; this estimate is found below to be in good agreement with the experimental data.


\section{Rigorous analysis}\label{sec:methodology}
Now, having understood the expected qualitative behavior of the photothermal nonlinearity in plasmon-assisted photocatalysis, we turn to describe the methodology employed to study it {\XYZ rigorously}. 
{\XYZ In order to properly account for the non-trivial reactor geometry and the multitude of materials, we used a numerical software package (COMSOL Multiphysics) to solve the time-independent heat (or Poisson) equation with temperature-dependent parameters. 

Two generic configurations were used in plasmon-assisted photocatalysis experiments. One is based on pellet geometries (Fig.~\ref{fig:scheme}(a)) and the other on NPs in a liquid suspension (Fig.~\ref{fig:scheme}(b)). In both cases, we distinguish between the thermal conductivities of the catalyst sample and of its surrounding to account for the inhomogeneity of the thermal properties, namely, we solve}
\begin{align}
\begin{cases}
\nabla \cdot \left[\kappa_\textrm{cata}(T({\bf r})) \nabla T({\bf r})\right] = - p_\textrm{abs}({\bf r}),&\textrm{inside the catalyst,} \\
\nabla \cdot \left[\kappa_\textrm{holder} (T({\bf r}))\nabla T({\bf r})\right] = 0,&\textrm{in the sample holder},  \\
\nabla \cdot \left[\kappa_\textrm{gas} (T({\bf r}))\nabla T({\bf r})\right] = 0,&\textrm{elsewhere.}
\end{cases}\label{eq:heat-poisson-nlty}
\end{align}
{\XYZ Here, $\kappa_\textrm{cata}(T({\bf r}))$, $\kappa_{\textrm{holder}}(T({\bf r}))$ and $\kappa_{\textrm{gas}}(T({\bf r}))$ are the temperature-dependent thermal conductivities of the catalyst~\cite{Y2-eppur-si-riscalda,Un-Sivan-sensitivity}, sample holder, and surrounding gas, respectively}. Finally, $p_\textrm{abs}$ is the heat source density induced by the light absorption in the random metal NP array; it is well described by the effective medium approximation for the electromagnetic properties of the catalyst sample~\cite{Un-Sivan-sensitivity}. {\XYZ When the skin (penetration) depth is much smaller than the sample thickness one can neglect} the temperature variation within the skin depth, namely,
\begin{align}\label{eq:pabs}
p_\textrm{abs}({\bf r}) = \int  \dfrac{i_\textrm{inc}(\rho,\omega)}{\delta_\textrm{skin}(\omega,T_1)}\exp\left(-\dfrac{z}{\delta_\textrm{skin}(\omega,T_1)}\right) d\omega.
\end{align}
Here, $\rho$ is the distance from the propagation optical axis, $z$ is the distance along the propagation direction of the incident beam from the top surface of the catalyst sample, and $i_\textrm{inc}(\rho,\omega)$ describes the transverse spatial and spectral profile of the incident beam. $T_1$ represents the temperature of the top layer of the catalyst sample and is now an {\it unknown} variable which needs to be determined by solving Eq.~\eqref{eq:heat-poisson-nlty} and Eq.~\eqref{eq:pabs} self-consistently. 

In practice, in Eq.~(\ref{eq:pabs}) (specifically, when calculating the absorption cross-section), $T_1$ is chosen to be 300 K since the only available data for the metals used as catalysts in the papers we analyze below (Ru and Cu) are at 300 K. In fact, this is a very good approximation for Eq.~\eqref{eq:pabs} because, as pointed out in Section~\ref{sec:qualitative}, the temperature dependence of the permittivities has a minor effect on the overall photothermal response when the skin (penetration) depth is much smaller than the sample thickness. In addition, this approximation also allows significant time and computational resource-saving. In contrast, when the skin depth is comparable to the sample thickness, one needs to account for the temperature dependence of the permittivities and the temperature variation within the skin depth (see Supplemental Information Section~\ref{suppsec:T_dep_derivation} for a complete derivation). In this case, the expression for the heat source density becomes much more complicated than Eq.~\eqref{eq:pabs} (see Eq.~\eqref{suppeq:pabs_closeform}) and one needs to solve Eq.~\eqref{eq:heat-poisson-nlty} with Eq.~\eqref{suppeq:pabs_closeform} self-consistently. {\XYZ The earlier analysis of the temperature dependence of the permittivities~\cite{Sivan-Chu-high-T-nl-plasmonics,Gurwich-Sivan-CW-nlty-metal_NP,IWU-Sivan-CW-nlty-metal_NP} imply that even in this case one should expect a sublinear growth of the temperature. }

{\XYZ In order to limit the simulation domain to a realistic and significant volume, we set a convection heat flux boundary condition at the outer surfaces of the {\XYZ simulation domain}; this models the heat transfer driven by the temperature difference between the {\XYZ simulation domain} and the distant surrounding (assumed to be at 20$^\circ$C). The associated heat transfer coefficient (denoted by $h$ in COMSOL Multiphysics) is used as an adjustable parameter to fit the experimental results in the small intensity limit for which one can neglect the temperature dependence of the thermal conductivities. In this case, Eq.~\eqref{eq:heat-poisson-nlty} becomes a linear Poisson equation, the solution of which is the linear approximation of the solution of Eq.~\eqref{eq:heat-poisson-nlty}. The obtained value for $h$ has a negligible effect on the photothermal nonlinearity (see Fig.~\ref{suppfig:JieLiu_NanoLett_2019_temp_map_T1_T2_h_para}). We emphasize that the frequently employed fixed temperature boundary condition does not represent correctly the heat transfer from the reactor to the environment. Instead, it effectively brings the distant cool regions closer to the reactor, and thus, cools the reactor in a manner that is inconsistent with measured experimental data. }

For simplicity, we apply the non-convection approximation. This was justified by simplistic estimates~\cite{Y2-eppur-si-riscalda} which showed that the standard gas flow level used is not expected to be significant in removing the generated heat, and further supported by simulations of natural and forced convection done in Ref.~\citenum{IW-Sivan-redox-paper}; this effect anyhow is not expected to affect the nonlinear response.

\section{Analysis of Experiments}\label{sec:experiments}
{\XYZ In the following, we apply the model described above to two representative experiments. Both involved pellet geometries which are typically composed of a large number ($\sim 10^{12} - 10^{14}$) of few nm metal NPs randomly distributed within a highly sparse 3D powder} of micron-size metal oxide particles, see Fig.~\ref{fig:scheme}(a); gases occupy the volume between the NPs. In a typical reaction chamber, the catalyst sample (typically a few mm in size) is placed on a (stainless steel) sample holder and the reaction rate is then measured under a specific illumination and/or under resistive heating.

The heat conduction in such catalyst sample might be expected to be dominated by the solids (the sample holder and the oxide) since their thermal conductivities are much larger than that of the gases. However, this is not the case and instead, as shown previously~\cite{Un-Sivan-sensitivity}, it is the thermal conductivity of the gases that dominates. To see that, we first note that since {\XYZ for all practically useful samples and in all the cases we study below,} the heat generation occurs primarily on the top layer of the catalyst sample, i.e., away from most of the sample holder, the sample holder is effective in reducing the bottom-surface temperature but is less effective in reducing the top-surface temperature (see Supplemental Information Section~\ref{suppsec:thermal_simulations} for details). 

Second, in such samples, the oxide fill factor typically only reaches $\sim 10$\% (see Eq.~\eqref{suppeq:volume_fraction}). Indeed, the sample consists of a highly sparse random array of micron size oxide particles, which are barely touching each other, see Fig.~\ref{fig:scheme}(a); thus, somewhat unintuitivelythe gas serves as the bottle-neck for the heat conduction in the catalyst sample and the heat conduction through the oxide is highly inefficient. This conceptual picture has been verified by a hierarchy of effective medium approaches which was found to be in good agreement with experimental results (see Ref.~\citenum{kappa_powders}). Specifically, that work showed that the Maxwell-Garnett model~\cite{birdtransport,Pietrak-eff-kappa-compos-2014} provides an excellent approximation for the effective thermal conductivity $\kappa^\textrm{eff}_\textrm{cata}$ of the catalyst sample we analyze below, namely (see Supplemental Information Section~\ref{suppsec:thermal_simulations} for details),
\begin{align}\label{eq:kappa_cata_eff}
\kappa^\textrm{eff}_\textrm{cata} = \kappa_\textrm{gas} + \dfrac{3 f_\textrm{oxide}\kappa_\textrm{gas}}{\dfrac{\kappa_\textrm{oxide} + 2\kappa_\textrm{gas}}{\kappa_\textrm{oxide} - \kappa_\textrm{gas}} - f_\textrm{oxide}},
\end{align}
where $\kappa_\textrm{oxide}$ is the thermal conductivity of the oxide. Thus, since the fill factor of the oxide is small ($f_\textrm{oxide} \ll 1$), the effective thermal conductivity of the catalyst sample becomes approximately $\kappa^\textrm{eff}_\textrm{cata} \approx (1 + 3 f_\textrm{oxide})\kappa_\textrm{gas}$ (see Eqs.~\eqref{suppeq:kappa_cata}-\eqref{suppeq:kappa_cata_approx}), i.e., it is very close to that of the gas mixture (see Fig.~\ref{suppfig:kappa_cata_pellet}). Due to the above, in the examples below we set $\kappa_{cata} = \kappa_{cata}^{eff}$.

\subsection{Analysis of experiments from Li {\em et al}., Nano Letters {\bf 2019}, 19, 1706-1711~\cite{Liu-Everitt-Nano-Letters-2019}}\label{sec:JLiu_nanolett}
First, we look at the experimental results of Li {\it et al.}~\cite{Liu-Everitt-Nano-Letters-2019} who studied ammonia synthesis using a cesium-promoted, magnesium-oxide supported, ruthenium (Ru-Cs/MgO) catalyst. The Ru NPs in this study were estimated to be $\sim 2$nm in diameter. The catalyst sample (3mm height and 6mm diameter) was put in a reaction chamber equipped with a quartz window (which allows the catalyst sample to be illuminated at varied intensity and wavelength) and a temperature controller (which is used to heat up the catalyst sample). The illumination spot was set to have the same size as the catalyst surface area. Two thin thermocouples were inserted into the catalyst sample, one to measure the top-surface temperature (denoted by $T_1$) and the other to measure the bottom-surface temperature (denoted by $T_2$). A mixture of N\textsubscript{2}, H\textsubscript{2}, and Ar with a ratio of 1:3 for N\textsubscript{2}/H\textsubscript{2} flowed into the reactor at a total flow rate of 75 sccm. The gaseous product (NH\textsubscript{3}) was monitored by an online mass spectrometer. For these parameters, an estimate similar to the one done in Ref.~\citenum{Y2-eppur-si-riscalda} shows that the gas flow may affect the temperature by no more than 10\%, justifying the use of the non-convection approximation in the following analysis.

\begin{figure*}[h]
\centering
\includegraphics[width=0.7\textwidth]{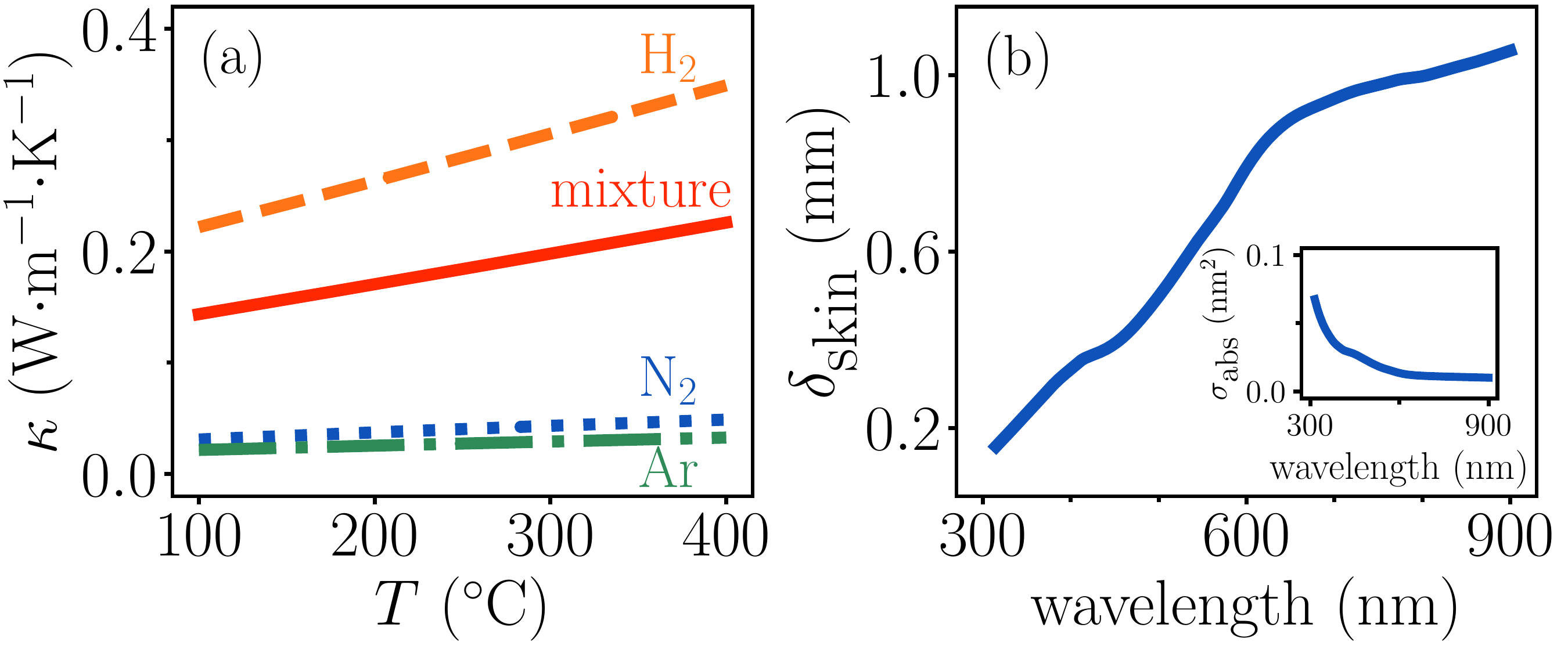}
\caption{(Color online) (a) Temperature dependence of the thermal conductivities of N\textsubscript{2}~\cite{kappa_N2_EngToolBox} (blue dotted line), H\textsubscript{2}~\cite{kappa_H2_EngToolBox} (orange dashed line), Argon~\cite{kappa_Ar_Saxena_Chen_1975} (green dash-dotted line) and the mixture used in Ref.~\citenum{Liu-Everitt-Nano-Letters-2019} (red solid line). (b) Light penetration (skin) depth and the absorption cross-section of the Ru NP (inset) as a function of the illumination wavelength.}
\label{fig:JieLiu_NanoLett_2019_k_dsk}
\end{figure*}

The temperature-dependent thermal conductivities of the input gases~\cite{kappa_N2_EngToolBox,kappa_H2_EngToolBox,kappa_Ar_Saxena_Chen_1975} are shown in Fig.~\ref{fig:JieLiu_NanoLett_2019_k_dsk}(a). Since H\textsubscript{2} has a small molecular mass and a small molecule size, its thermal conductivity is much larger than that of N\textsubscript{2} and Ar. As a result, the thermal conductivity of the gas mixture is $\sim 60\%$ of the H\textsubscript{2} thermal conductivity. The thermal conductivity of the catalyst sample is related to the volume fraction of the oxide using the Maxwell Garnett equation~\cite{birdtransport,Pietrak-eff-kappa-compos-2014,Y2-eppur-si-riscalda} (see Supplemental Information Section~\ref{suppsec:thermal_simulations}).

The absorption cross-section of the Ru NPs and the light penetration depth in the sample is calculated using the permittivity of Ru at 300K from Ref.~\citenum{Adachi_book_optical_const_metal} and Eq.~(\ref{eq:skin_depth}), see Fig.~\ref{fig:JieLiu_NanoLett_2019_k_dsk}(b). For wavelengths $300 < \lambda < 600$nm, the penetration (skin) depth (Eq.~\eqref{eq:skin_depth}) is $\delta_{\textrm{skin}} < 750 \mu$m, indeed much smaller than the sample thickness; thus, by Ref.~\citenum{Un-Sivan-sensitivity}, we do not expect the numerical results to be sensitive to the exact parameters in this spectral regime. However, for $\lambda > 700$nm, we find that $\delta_{\textrm{skin}}\gtrsim$ 1 mm, so that here the penetration (skin) depth is only $2-3$ times thinner than the sample thickness, see Fig.~\ref{fig:JieLiu_NanoLett_2019_k_dsk}(b); accordingly, one may expect a slight sensitivity to the various parameters in this regime. In addition, we assume that the transverse profile of the illumination intensity is uniform (denoted as $I_\textrm{inc}$).

\begin{figure*}[h]
\centering
\includegraphics[width=1\textwidth]{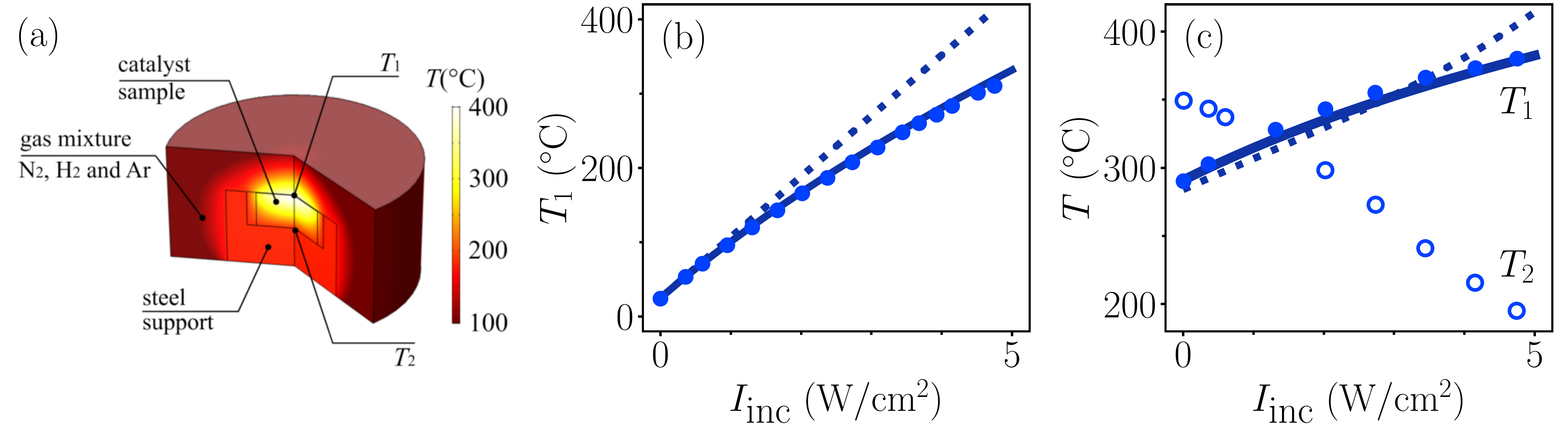}
\caption{(Color online) (a) Details of the photocatalytic chamber and the temperature distribution at the illumination intensity $I_{\textrm{inc}} = 5$W/cm\textsuperscript{2}. (b) $T_1$ (blue solid symbols: experimental data extracted from Ref.~\citenum{Liu-Everitt-Nano-Letters-2019}, blue solid line: COMSOL simulation, blue dotted line: COMSOL simulation with the linear approximation) as a function of the illumination intensity for the blue light source without resistive heating.
(c) Same as (b) but under resistive heating. The blue opened symbols represent the experimental data of $T_2$.
}\label{fig:JieLiu_NanoLett_2019_temp_map_T1_T2}
\end{figure*}

To obtain the temperature distribution, we perform a full simulation including the catalyst sample, the sample holder, and the reaction chamber with a few simplifications. Specifically, we simplified the complicated reaction chamber, by assuming it has a cylindrical shape (1 cm height and 2 cm diameter) and that the catalyst sample (having the same size as used in the experiment) is placed on a steel holder, see Fig.~\ref{fig:JieLiu_NanoLett_2019_temp_map_T1_T2}(a)~\footnote{The geometry of the reactor chamber, of the sample holder and of the catalyst sample were directly obtained from the authors of Refs.~\citenum{Liu-Everitt-Nano-Letters-2019,Liu_thermal_vs_nonthermal}.}. 

In Ref.~\citenum{Liu-Everitt-Nano-Letters-2019}, two experiments were performed. In the first, {\XYZ no resistance-heating was used}; the simulation results of the top surface temperature $T_1$ for $h \approx 70$ W/(m$^2\cdot$K) demonstrate an excellent match to the experimental data, see Fig.~\ref{fig:JieLiu_NanoLett_2019_temp_map_T1_T2}(b) (the simulation result of the bottom surface temperature $T_2$ is shown in Fig.~\ref{suppfig:JieLiu_NanoLett_2019_temp_map_T1_T2}(a)). In order to demonstrate the actual level of the photothermal nonlinearity, we also perform a simulation under the linear approximation, namely, we neglect the temperature dependence of the thermal conductivity of the gas mixture. Fig.~\ref{fig:JieLiu_NanoLett_2019_temp_map_T1_T2}(b) shows that the linear approximation overestimates the nonlinear solution by more than $30\%$ for a temperature rise of $\sim 300^\circ$C; this is an unusually large nonlinearity. 

In the second experiment, the authors studied the effect of the temperature gradient on the reaction rate and the dependence of the temperature gradient on the illumination wavelength and intensity. To do that, they measured the intensity-dependent $T_1$, $T_2$ and reaction rate using four different light sources (UV, blue light, white light and NIR) and adjusted the resistive heating such that the equivalent temperature\footnote{The equivalent temperature $T_{\textrm{e}}$ is defined in Refs.~\citenum{Liu_thermal_vs_nonthermal,Liu-Everitt-Nano-Letters-2019} through the relation $e^{-E_{\textrm{a}}/k_{\textrm{B}} T_{\textrm{e}}} = \dfrac{1}{T_2 - T_1}\int_{T_1}^{T_2} e^{-E_{\textrm{a}}/k_{\textrm{B}} T} dT$.} remains the same (325$^\circ$C) for all four light sources and all intensities~\cite{Liu-Everitt-Nano-Letters-2019}. Accordingly, we use the same simulation configuration to calculate the temperature distribution. Since the description of the resistive heating apparatus is not available in Ref.~\citenum{Liu-Everitt-Nano-Letters-2019}, we simply set the temperature of the bottom of the catalyst sample to the reported $T_2$ as a constraint so as to mimic the resistive heating. The simulation results for the top surface temperature $T_1$ again demonstrate an excellent match with the experimental data for blue, UV, and white light sources\footnote{The simulation results for UV and white light sources are shown in Fig.~\ref{suppfig:JieLiu_NanoLett_2019_temp_map_T1_T2}(b) because they almost overlap with the results for the blue light source. This is a direct consequence of the weak sensitivity of the temperature distribution to the illumination wavelength when the penetration depth is much smaller than the sample thickness, conforming with the analysis in Ref.~\citenum{Un-Sivan-sensitivity}.}, see Fig.~\ref{fig:JieLiu_NanoLett_2019_temp_map_T1_T2}(c) and Fig.~\ref{suppfig:JieLiu_NanoLett_2019_temp_map_T1_T2}(b). We find that the linear approximation deviates from the nonlinear solution by $\sim 10$\% for a temperature rise of $\sim 100^\circ$C. For the NIR light source, the simulation results fit well the experimental data, see Fig.~\ref{suppfig:JieLiu_NanoLett_2019_temp_map_T1_T2}(b). As mentioned, in this case, the light penetration depth is closer to the sample thickness (see Fig.~\ref{fig:JieLiu_NanoLett_2019_k_dsk}(b)) so that the temperature distribution might become less insensitive to the change of permittivity of the Ru NPs with temperature. In that regard, the use of the Ru permittivity data at $300$K in the simulation is one possible reason for the small mismatch between the simulation results and the experimental data. Except for this minor discrepancy, the analysis here reinforces the conclusion of Ref.~\citenum{Liu-Everitt-Nano-Letters-2019} that the catalytic effect of the Ru-Cs/MgO system on the ammonia synthesis reaction is purely thermal.

An earlier work~\cite{Liu_thermal_vs_nonthermal} by the same group employed a similar experimental system to study the carbon dioxide hydrogenation reaction using a titanium oxide supported Rhodium catalyst. Although the gas composition was different from that in Ref.~\citenum{Liu-Everitt-Nano-Letters-2019}, the thermal conductivities of the gas mixtures~\cite{kappa_CO2_EngToolBox,kappa_H2_EngToolBox,kappa_Ar_Saxena_Chen_1975} were similar since the gas mixtures had a similar fraction of H\textsubscript{2} in these two works. We simulated the temperature for the experiment without resistive heating at low temperatures (25$^\circ$C $< T_1 < 120^\circ$C) and found very good agreement with the experimental data (see Supplemental Information Fig.~\ref{suppfig:JieLiu_NanoLett_2018_T1_T2}). This explains the observed photothermal nonlinearity ($\sim$ 10\%) for a similar temperature rise of $\sim 100^\circ$C, see Fig.~S8 in Ref.~\citenum{Liu_thermal_vs_nonthermal}. However, a much larger nonlinearity was reported in the experiments with resistive heating at high temperatures ($250^\circ$C $< T_1 < 450^\circ$C); in fact, some of the data shows an unusually large increase in $T_1$ for $300^\circ$C $\lesssim T_1 \lesssim 375^\circ$C and then a small growth rate in $T_1$ for $T_1 \gtrsim 375^\circ$C. Accordingly, it cannot be explained by our (nonlinear) thermal model. Ruling out possible measurement artifacts in the data, it is natural to advocate for the possibility of non-thermal electrons contributing to the catalyzed reaction rate. However, a convincing interpretation of this experiment might require a dedicated explanation of the origin of the large nonlinearity.

\subsection{Analysis of experiments from Zhou {\em et al.}, Science, {\bf 2018}, 362, 69~\cite{Halas_Science_2018}}

\begin{figure*}[h]
\centering
\includegraphics[width=1\textwidth]{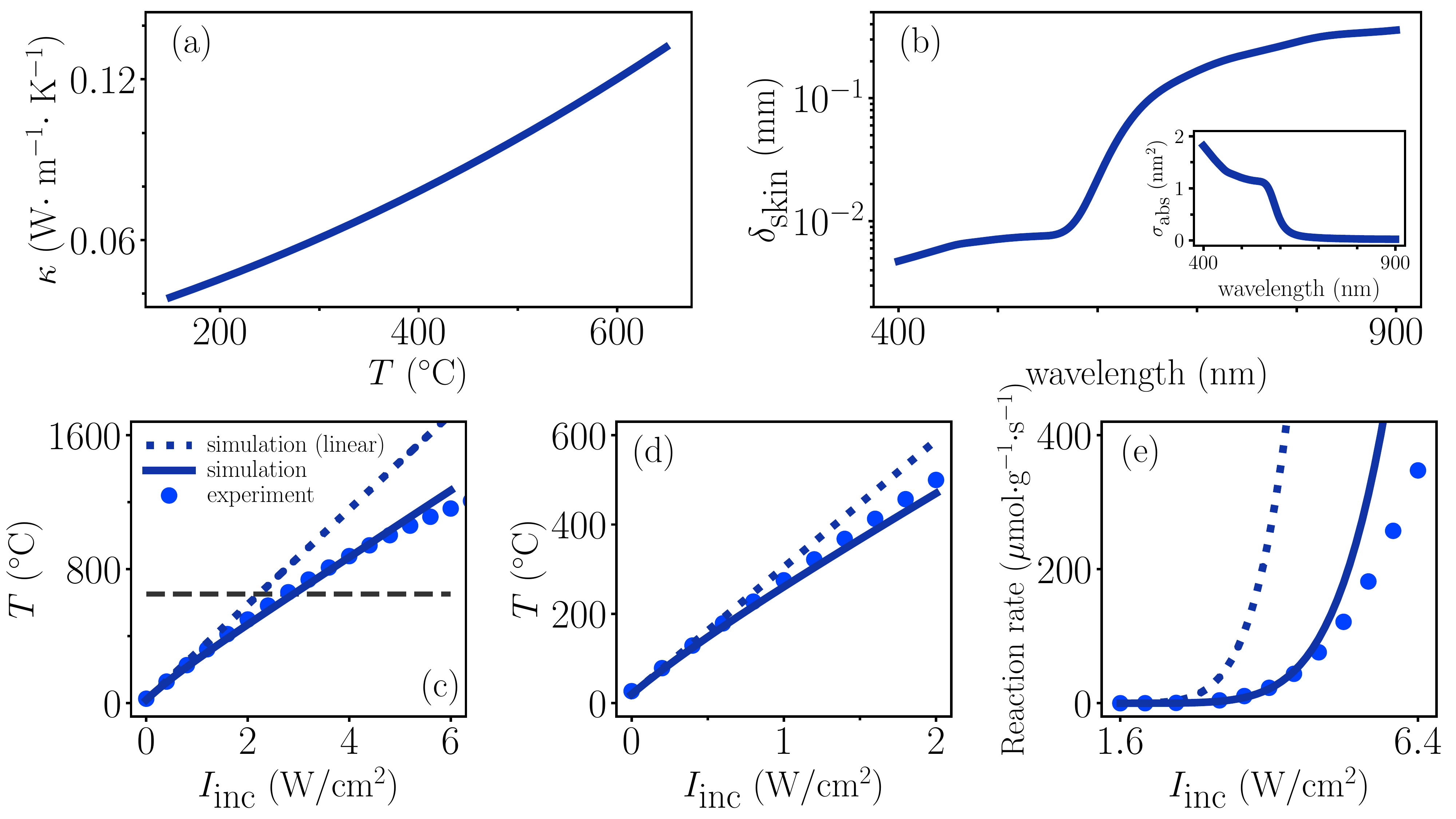}
\caption{(Color online) (a) Temperature dependence of the thermal conductivity of NH\textsubscript{3}~\cite{Afshar_kappa_NH3,kappa_NH3_EngToolBox}. (b) Light penetration depth and the absorption cross-section of the Cu-Ru NP (inset) as a function of the wavelength. (c) Simulation results of the temperature at the center of the top surface as a function of the illumination intensity (blue solid line) and its linear approximation (blue dotted line); The black dashed line represents the upper bound of the temperature range of the NH\textsubscript{3} thermal conductivity measured in Ref.~\citenum{Afshar_kappa_NH3}. (d) A zoom-in of (c) at the low illumination region. (e) Calculated reaction rates (the blue solid line and the blue dotted line) as a function of the illumination intensity. The blue circles in (c), (d) and (e) represent the numerical fit of the experimental data using the temperature-shifted Arrhenius law. } \label{fig:Halas_Science_2018_k_dsk_temp}
\end{figure*}

We now move on to show that the photothermal nonlinearity also explains the experimental measurements described in Ref.~\citenum{Halas_Science_2018}. This work employed a similar setup to study ammonia decomposition on a MgO-Al\textsubscript{2}O\textsubscript{3} supported Cu-Ru catalyst, however, unlike the papers studied in Section~\ref{sec:JLiu_nanolett}, {\XYZ the nonlinearity is much stronger due to the higher temperature rises caused by the much smaller thermal conductivity of NH\textsubscript{3}~\cite{Afshar_kappa_NH3,kappa_NH3_EngToolBox} (see Fig.~\ref{fig:Halas_Science_2018_k_dsk_temp}(a)), as well as the higher incident intensities used (see Fig.~\ref{fig:Halas_Science_2018_k_dsk_temp}(c))}. More importantly, while the authors of~\cite{Liu-Everitt-Nano-Letters-2019} measured the sample temperature properly, significant concerns regarding the validity of the temperature measurements in~\cite{Halas_Science_2018} have been raised (see discussion in Refs.~\citenum{anti-Halas-Science-paper,Y2-eppur-si-riscalda,R2R,Dubi-Sivan-APL-Perspective,anti-Halas-NatCat-paper}). Therefore, in what follows, we rely only on a {\em calculation} of the temperature as well as on an extraction of it from the reaction rate using the Arrhenius equation. 
These approaches were shown to match well with the fitted temperatures, see Refs.~\citenum{anti-Halas-Science-paper,Y2-eppur-si-riscalda,anti-Halas-NatCat-paper}.

We follow the procedure described in Ref.~\citenum{Y2-eppur-si-riscalda} to extract the temperature out of the reaction rate. Specifically, we first obtain the activation energy 1.18eV by fitting the experimental data in the dark (Fig.~1d in Ref.~\citenum{Halas_Science_2018}) to an Arrhenius curve. Next, due to the failure to measure the light-induced temperature rise in Ref.~\citenum{Halas_Science_2018} (see~\citenum{anti-Halas-Science-paper,R2R}), we distinguish between the actual temperature of the reactor (denoted as $T(I_\textrm{inc})$) and the experimentally-measured temperature (denoted as $T_{\textrm{M}}$, see Fig.~S11D in Ref.~\citenum{Halas_Science_2018}). The difference between $T(I_\textrm{inc})$ and $T_{\textrm{M}}$ was found to be $T(I_\textrm{inc}) = T_{\textrm{M}} + \tilde{a} I_\textrm{inc} + \tilde{b} I_\textrm{inc}^2$, where $I_\textrm{inc}$ is in W/cm\textsuperscript{2}, $T(I_\textrm{inc})$ and $T_{\textrm{M}}$ are in K. The experimentally-measured temperature $T_{\textrm{M}}$ vs. $I_\textrm{inc}$ is also fitted to a second-order polynomial, giving $T_\textrm{M} = 298 + 80 I_\textrm{inc} - 3.8 I_\textrm{inc}^2$. Then, we fit the measured reaction rate data under illumination (Fig.~1d in Ref.~\citenum{Halas_Science_2018}) to a temperature-shifted Arrhenius curve~\cite{Y2-eppur-si-riscalda}, i.e., $R(I_\textrm{inc}) = R_0 \exp\left(-\dfrac{E_\textrm{a}}{k_B T(I_\textrm{inc})}\right)$, leading to $T(I_\textrm{inc}) = T_\textrm{M} + 180 I_\textrm{inc} - 8 I_\textrm{inc}^2$, see Figs.~\ref{fig:Halas_Science_2018_k_dsk_temp}(c), (d) and (e). This shows that the photothermal nonlinearity becomes nearly $\sim 50\%$ at the highest intensity used. This also shows that the temperature rise due to photon absorption becomes a sublinear function of the illumination intensity at high temperatures (as predicted in Refs.~\citenum{Sivan-Chu-high-T-nl-plasmonics,Gurwich-Sivan-CW-nlty-metal_NP,IWU-Sivan-CW-nlty-metal_NP} for a single metal nanoparticle) so that unlike the claims in Ref.~\citenum{thm_hot_e_faraday_discuss_2019} (see p. 270 and on), the maximal temperature reached is $\sim 1400^\circ$C rather than $\sim 2700 ^\circ$C. This result shows that claims that our thermal model predicts temperatures which are unrealistically high are \textit{simply}  incorrect.

The question remains - what is the reason for this massive sublinearity in the temperature-rise-versus-intensity? To answer this question, we adapt the simulation configuration used in Section~\ref{sec:JLiu_nanolett} to the experimental setup described in Ref.~\citenum{Halas_Science_2018} and simulate the temperature distribution via the heat transfer module of COMSOL Multiphysics. The thermal conductivity of NH\textsubscript{3} is taken from the experimental measurement of Refs.~\citenum{Afshar_kappa_NH3,kappa_NH3_EngToolBox} for the temperature range of 52 - 652$^\circ$C~\footnote{Although the thermal conductivity of NH\textsubscript{3} was measured at the pressures of 12.9, 26.5, and 45.0 kN/m\textsuperscript{2} (0.127, 0.262 and 0.444 atm) in Ref.~\citenum{Afshar_kappa_NH3}, this data can be used for the simulation at 1 atm since the gas thermal conductivity is very weakly-dependent on the pressure~\cite{Francis-thermodynamics-kinetic-theory-book}. To justify this, we compared the data from Ref.~\citenum{Afshar_kappa_NH3} with the data measured at 1 atm but in a lower temperature range ($-33.6$ - $426.9^\circ$C) provided in Ref.~\citenum{kappa_NH3_EngToolBox} and found good agreement between these two sets of data.}. The measured data were also fitted in Ref.~\citenum{Afshar_kappa_NH3} by the following cubic polynomial, $\kappa_{\textrm{NH\textsubscript{3}}}(T) = 5.237\times10^{-4} + 5.179\times 10^{-5} T + 8.404\times10^{-8}T^2+1.557\times 10^{-11}T^3$, where $\kappa_{\textrm{NH\textsubscript{3}}}$ is in W/(m$\cdot$K) and $T$ is in K, as shown in Fig.~\ref{fig:Halas_Science_2018_k_dsk_temp}(a). This cubic polynomial is used in our simulations to extrapolate the NH\textsubscript{3} thermal conductivity for temperatures higher than 652$^\circ$C.

The catalyst sample was illuminated by a pulsed broadband white-light source (Fianium, WL-SC-400-8, 400-900 nm, pulse duration 4 ps, repetition rate 80 MHz, and a 2 mm diameter beam profile on the catalyst sample surface) without applying any resistive heating~\cite{Halas_Science_2018}. The penetration (skin) depth to the sample is calculated using the permittivity data of Cu and Ru at 300K~\cite{Adachi_book_optical_const_metal}. We find that the penetration depth is much smaller than the sample thickness, see Fig.~\ref{fig:Halas_Science_2018_k_dsk_temp}(b). In addition to the temperature distribution, we also calculated the reaction rate based on the Arrhenius equation using the calculated temperature at the center of the top sample surface. Our simulation results show satisfactory agreement with the fitted data, although they are independent of the fit procedure. In fact, the agreement extends up to 1000$^\circ$C, i.e., even beyond the expected bound for the validity of the used values for the thermal conductivity of the host, see Figs.~\ref{fig:Halas_Science_2018_k_dsk_temp}(c) and (d). In particular, the temperature at the center of the top surface increases monotonically with the illumination intensity but with a decreasing slope, see Fig.~\ref{fig:Halas_Science_2018_k_dsk_temp}(c). In the original manuscript~\cite{Halas_Science_2018}, this sublinear growth of the temperature rise with the illumination intensity was {\em incorrectly} attributed to the temperature dependence of the thermal conductivity of the oxide support. This claim is, however, invalid for two reasons. First, this claim would lead to an opposite trend to that observed since the thermal conductivity of the MgO-Al\textsubscript{2}O\textsubscript{3} support {\em decreases} with temperature~\cite{Hofmeister_kappa_MgO_Al2O3}. Second, the temperature dependence of the thermal conductivity of the oxide support, in fact, was shown in Ref.~\citenum{Un-Sivan-sensitivity} to have a negligible effect on the overall nonlinearity because of the small volume fraction of oxides in the catalyst sample (using the sample mass and the sample volume reported in Ref.~\citenum{Halas_Science_2018}). Instead, in agreement with the qualitative analysis and the modelling of the papers studied in Section~\ref{sec:JLiu_nanolett}, the sublinear growth of the temperature rise with the illumination intensity is mainly due to the increase of the gas thermal conductivity with temperature~\cite{Afshar_kappa_NH3,kappa_NH3_EngToolBox}. 

One of the possible reasons for the slight discrepancy observed for $T > 1000^\circ$C between the simulation results and the fitted data at high intensities is the inaccuracy induced by the extrapolation to the thermal conductivity data of NH\textsubscript{3}. This is because the third-order term ($T^3$) in the cubic polynomial used for the extrapolation has the same order of magnitude as the second-order term ($T^2$) for temperatures higher than 1000$^\circ$C, i.e., it is likely that higher-order terms are required for better accuracy. Another possible but minor reason is the usage of the permittivity data at 300K for Cu in the simulation\footnote{Only limited data is available, e.g., in Ref.~\citenum{Pells-Shiga}, only the imaginary part of the permittivity for photon energy 1.8 - 6eV at 5 different temperatures (77K, 295K, 575K, 770K, and 920K) are provided; in Ref.~\citenum{Johnson_Christy_Cu_Ni_temp_dep_eps}, the permittivity data are provided at only 3 different temperatures (78K, 293K, and 423K).}. However, since the light penetration depth is quite smaller than the sample thickness, the change of the optical properties of Cu could only have a minor effect on the overall photothermal nonlinearity, see discussion in Ref.~\citenum{Un-Sivan-sensitivity}. Since comprehensive data of the temperature dependence of the metal permittivity and of the gas thermal conductivity hardly exist at such high-temperature regimes, the resolution of this discrepancy requires further experimental study. Nevertheless, our simulation result shows that the photothermal nonlinearity plays a non-negligible role in reducing the growth rate of the rising temperature and, hence, of the photocatalysis reaction rate.

\section{Discussion}\label{sec:discussion}
The analysis presented above was based on the initial modeling  of the low-temperature response. Once the unknown system parameters were determined by fit to the experimental data, the nonlinear photothermal response observed experimentally was modeled accurately using the known temperature dependence of the various material constituents, i.e., using no additional fit parameters. The success of the photothermal analysis shows that as qualitatively predicted in Section~\ref{sec:qualitative}, the nonlinear response in the temperature rise originates from the temperature dependence of the (effective) thermal conductivity of the host, and not from the response of the metal itself.

This result also confirms the error in previous claims on a different source of nonlinearity for these systems, and shows that claims that the thermal model leads to unrealistic high temperatures are simply incorrect, see discussion in Ref.~\citenum{R2R}. The metal nonlinearity could be of significance only for very thin plasmonic catalysts or even on the single nanoparticle level (see discussion in Refs.~\citenum{Sivan-Chu-high-T-nl-plasmonics,Gurwich-Sivan-CW-nlty-metal_NP,IWU-Sivan-CW-nlty-metal_NP}). These systems are, however, usually of more fundamental rather than practical importance.

Unlike the generally weak effect of ``hot'' electrons, the photothermal nonlinearity is very strong. Therefore, this effect must be quantified before any claim for ``hot'' electron action can become convincing, and should not be ignored even at low illumination intensities. In practice, the rather large uncertainty in the magnitude of the associated nonlinear response coefficients means that only ``hot'' electron effects which are clearly greater than this uncertainty can be deduced. Unfortunately, satisfying these conditions poses a severe constraint on claims for ``hot'' electron dominance.

While the effect of the rising temperature on the absorptivity of the metal NPs may have a negligible effect on the overall temperature distribution, it may have a significant effect on the thermal emissivity via the Kirchhoff Law of Radiation, an effect already demonstrated experimentally~\cite{Vasque-Thesis}. This is relevant for a correct determination of the temperature using thermal imaging at mid-IR frequencies, i.e., the change of emissivity at those frequencies would need to be accounted for at high temperatures. To the best of our knowledge, this was not done so far in the context of plasmon-assisted photocatalysis, see discussion in Ref.~\citenum{Baffou-Quidant-Baldi}.

At the high temperatures at which significant photothermal nonlinearity may be observed, the NPs themselves may undergo geometrical and morphological changes and eventually may even melt (at a temperature which may be significantly lower than the bulk melting temperature). This possibility was discussed in great detail in Ref.~\citenum{thm_hot_e_faraday_discuss_2019} p. 270 and on. Briefly, while this effect may be possible, it was not probed directly, and it is a-priori expected not to affect the temperature distribution so much, especially not when $\delta_\textrm{skin} \ll H$~\cite{Un-Sivan-sensitivity}. Therefore, although melting may occur, it does not affect the results discussed in the current work in a significant manner.

{\XYZ Lastly, we note that system similar to those studied here were extensively studied in the past in the context of composites with a high thermo-optical nonlinearity, see e.g., Refs.~\citenum{Smith_Boyd_EMT_kerr_media,Boyd-metal-nlty,Palpant_EMT_kerr_media,Khurgin_chi3,Donner_thermal_lensing} and in the context of various applications such as optical limiting~\cite{Blau_review_SA_RSA} or tunable optical devices~\cite{Donner_thermal_lensing,Quidant_plasmon_optofluid}. The main difference is that the particle density in those systems was typically much lower, such that some (even most) of the light was transmitted through the sample. In addition, the focus in these systems was on (nonlinear) changes of the optical response (permittivity, transmission etc.) rather than on the temperature rise (as above). 

Thermal effects in these systems were usually ignored, and the optical response was typically interpreted using a temporally- and spatially-local response. In that sense, it would be intriguing to study the thermal response in such systems, to see if (the temporally- and spatially-{\em non}local) thermal effects could explain some of the experimental observations, in particular, the strong dependence on the spatial~\cite{Maznev-2011,ICFO_Sivan_metal_diffusion,Sivan_Spector_metal_diffusion} and temporal extent of the illumination~\cite{Stoll_review,Langbein_PRB_2012,Biancalana_NJP_2012} and on the host properties~\cite{thermo-plasmonics-basics,Sivan-Chu-high-T-nl-plasmonics,Gurwich-Sivan-CW-nlty-metal_NP,IWU-Sivan-CW-nlty-metal_NP}. In these cases, changes to the host permittivity may be more important, because of macroscopic transmission changes and thermal lensing effects, which were the main motivation for these studies in the first place.

\section*{Author Contributions}
I.W. Un derived the formulation, performed the calculations, and wrote the paper. Y.D. performed the initial fits to the experimental data and together with Y.S. supervised the project and contributed to the writing of the paper.}


\section*{Conflicts of interest}
There are no conflicts to declare.

\section*{Acknowledgements}
I.W.U. and Y.S. were supported by Israel Science Foundation (ISF) grant (340/2020) and by Lower Saxony - Israel cooperation grant no. 76251-99-7/20 (ZN 3637).

\bibliographystyle{rsc}
\providecommand*{\mcitethebibliography}{\thebibliography}
\csname @ifundefined\endcsname{endmcitethebibliography}
{\let\endmcitethebibliography\endthebibliography}{}

\end{document}


\author{Ieng Wai Un,$^{a\ast}$, Yonatan Dubi$^{b}$ and Yonatan Sivan$^{a}$
\\
\normalsize{$^{a}$School of Electrical and Computer Engineering, Ben-Gurion University of the Negev, Israel}\\
\normalsize{$^{b}$Department of Chemistry, Ben-Gurion University of the Negev, Beer-Sheva, 8410501, Israel.}\\
\normalsize{$^\ast$E-mail: iengwai@post.bgu.ac.il}}

\maketitle

\newpage
\section{The energy absorbed density accounting for the temperature dependence of the permittivities}\label{suppsec:T_dep_derivation}
In Section~\ref{sec:methodology}, we have derived the absorbed power density using the temperature-independent permittivity approximation. In this section, we provide a complete derivation of the absorbed power density accounting for the temperature variation within the heat source and the temperature dependence of the permittivity. This is required when the skin (penetration) depth of the light in the sample is comparable to the sample thickness (for example, when the nanoparticle density is highly dilute, or when the wavelength of the illumination is far away from the resonance wavelength of the NPs). In this case, the heat source density becomes 
\begin{align}\label{eq:pabs_beer_lambert}
p_{\textrm{abs}}({\bf r}) = -\int i_{\textrm{inc}}(\rho,\omega)\dfrac{\partial \zeta(\rho, z,\omega)}{\partial z} d\omega,
\end{align}
where $\zeta(\rho,z,\omega)$ is an unknown longitudinal spatial profile of the incident illumination which needs to be determined by solving the Beer-Lambert equation for the absorption of light with a space-dependent penetration depth $\delta_{\textrm{skin}}(T({\bf r}),\omega)$ (elucidated below), namely, ${\partial\zeta(\rho,z,\omega)}/{\partial z} = -{\zeta(\rho,z,\omega)}/{\delta_{\textrm{skin}}(T({\bf r}),\omega)}$. This enables us to accommodate the temperature gradient build-up in the catalyst sample under the illumination. 

As before, we neglect the temperature non-uniformity within the individual NPs due to their small size and their high thermal conductivity, see justification in Refs.~\citenum{thermo-plasmonics-basics,Un-Sivan-size-thermal-effect}. Furthermore, we assume that the temperature varies slowly on a length scale of the illumination wavelength (see justification in Fig.~\ref{fig:scheme}(b)). These assumptions allow us to evaluate the temperature dependence of the absorption cross-section using a uniform metal permittivity and a uniform host permittivity, i.e., $\sigma_{\textrm{abs}}(\omega,T) = \sigma_{\textrm{abs}}(\varepsilon_m(\omega,T),\varepsilon_h(\omega,T),\omega)$. In this case, the absorption coefficient becomes
$$1/\delta_{\textrm{skin}}(T,\omega) = n_p\sigma_{\textrm{abs}}(\varepsilon_m(\omega,T),\varepsilon_h(\omega,T),\omega),$$ in analogy to Eq.~\eqref{eq:skin_depth}, and the absorbed power density has a closed form expression
\begin{align}\label{suppeq:pabs_closeform}
p_{\textrm{abs}}(\rho,z) = \int i_{\textrm{inc}}(\rho,\omega) n_p \sigma_{\textrm{abs}}(\omega,T(\rho,z))e^{- n_p \int_0^z \sigma_{\textrm{abs}}(\omega,T(\rho,z^{\prime})) dz^{\prime}} d\omega.
\end{align}
The temperature distribution is obtained by solving Eq.~\eqref{eq:heat-poisson-nlty} coupled with Eq.~\eqref{suppeq:pabs_closeform} self-consistently. 

\section{Supplemental simulation results for the experiments of Li \textit{et al.}~\cite{Liu-Everitt-Nano-Letters-2019}}
\begin{figure}[h]
\centering
\includegraphics[width=0.9\textwidth]{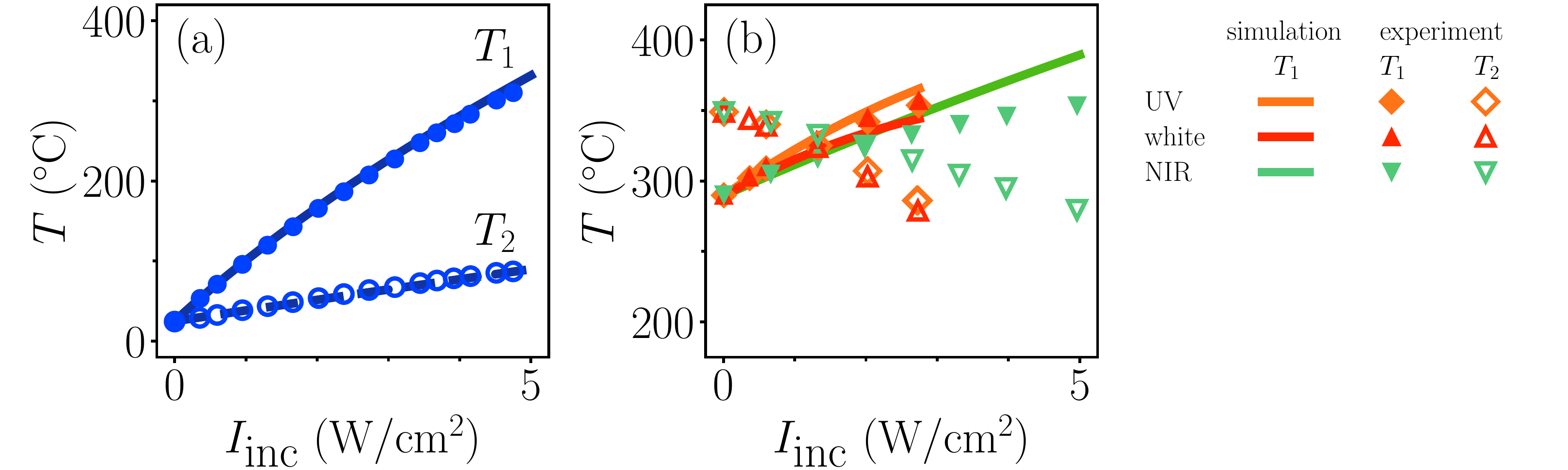}
\caption{(Color online) The top ($T_1$) and the bottom ($T_2$) surface temperatures (the simulation results are represented by the solid and the dashed lines; the experimental data are represented by the solid and the opened symbols.) as a function of the illumination intensity for the blue light source without resistive heating. (b) The same as (a) but under resistive heating. The orange, red and green lines (symbols) represents the simulation results (experimental data) for the UV, white and NIR light sources, respectively.}
\label{suppfig:JieLiu_NanoLett_2019_temp_map_T1_T2}
\end{figure}

{\XYZ Fig.~\ref{suppfig:JieLiu_NanoLett_2019_temp_map_T1_T2}(a) shows the simulation results of the top ($T_1$) and the bottom ($T_2$) surface temperatures for the case where the catalyst sample was illuminated using the blue light source (without resistive heating). One can see that the simulation results demonstrate an excellent match to the experimental data.  This indicates that our model is not only able to explain the photothermal nonlinearity observed in the experiment, but also can explain the large temperature difference built up by the illumination (see Section~\ref{suppsec:thermal_simulations} below). Fig.~\ref{suppfig:JieLiu_NanoLett_2019_temp_map_T1_T2}(b) shows the simulation results of the top ($T_1$) surface temperature for the cases where the catalyst sample was resistive-heated and was illuminated three different light sources (UV, white and NIR). 

}

\section{Electromagnetic and thermal simulations}
In the experiments~\cite{Liu-Everitt-Nano-Letters-2019,Liu_thermal_vs_nonthermal,Halas_Science_2018}, the catalyst pellet was a mixture of a highly sparse powder of metal nanoparticles and metal oxide microparticles; the gas mixture occupies the empty regions between the various particles. The catalyst pellet sits in a stainless steel sample holder and is put in a reaction chamber, see Fig.~\ref{fig:scheme}.

\subsection{Electromagnetic simulations}
In the electromagnetic simulations, the absorption of light is modeled using the effective medium approximation for the electromagnetic properties of the catalyst sample, see Section~\ref{sec:methodology} and Eq.~\eqref{suppeq:pabs_closeform}. The absorption cross-section of the NPs is calculated using Mie theory~\cite{Bohren-Huffman-book} and the bulk permittivity. This could be somewhat inaccurate for NPs of very small sizes (diameter $< 10$ nm) because the metal permittivity is different from the bulk permittivity due to the nonlocal effect~\cite{GdA_nonlocal_2008,GdA_nonlocal_2011}. However, for the conditions under which these experiments are conducted, the penetration depth is much shorter than the sample thickness (see Fig.~\ref{fig:JieLiu_NanoLett_2019_k_dsk}(b) and~\ref{fig:Halas_Science_2018_k_dsk_temp}(b)), so that effectively all light that enters the sample gets absorbed. Under these conditions, a change of the absorption cross-section of the NPs would only slightly modify the penetration depth, and will have essentially no effect on the overall temperature distribution in the sample (and hence, on the results in this work).

\subsection{Thermal simulations}\label{suppsec:thermal_simulations}
\begin{figure}[h]
\centering
\includegraphics[width=0.7\textwidth]{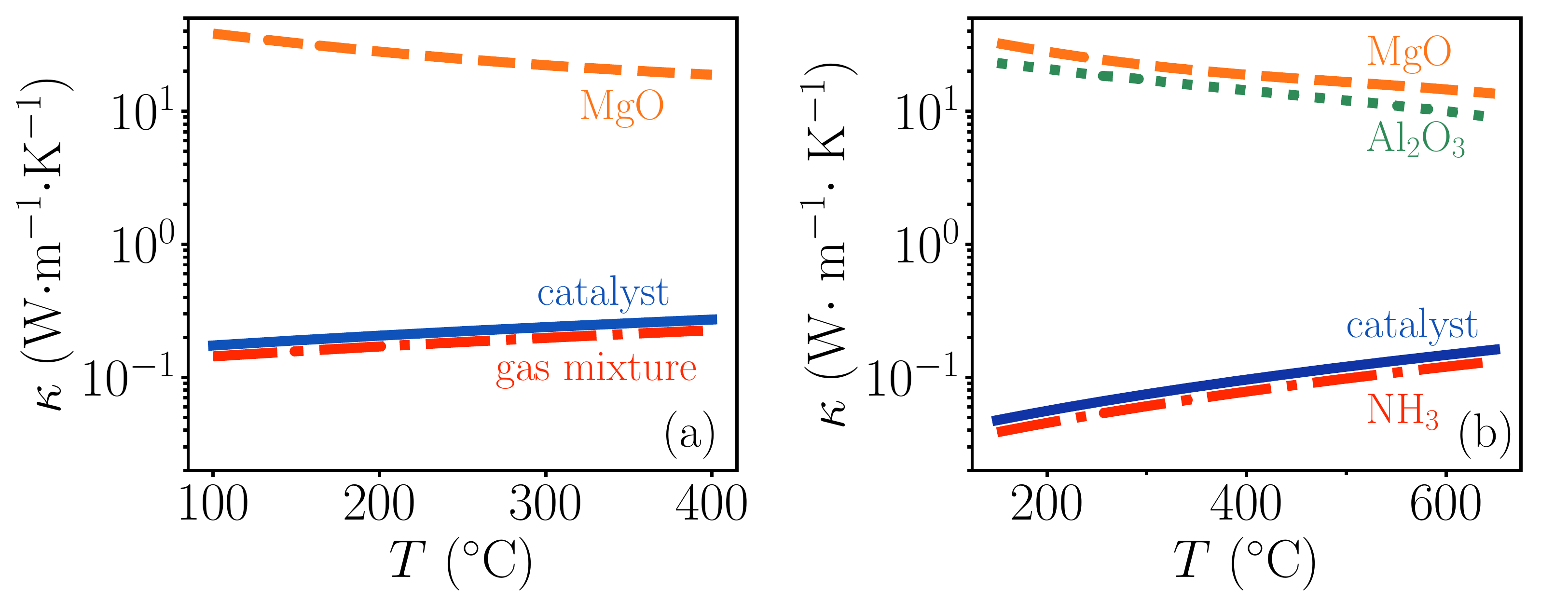}
\caption{(Color online) The temperature dependence of the thermal conductivities assigned to the simulations for (a) the experiment in~\cite{Liu-Everitt-Nano-Letters-2019} and (b) the experiment in~\cite{Halas_Science_2018}. }
\label{suppfig:kappa_cata_pellet}
\end{figure}

The temperature distribution is obtained from the steady-state solution of the heat equation Eq.~\eqref{eq:heat-poisson-nlty} with temperature-dependent coefficients ($\kappa^\textrm{eff}_\textrm{pellet}(T)$, $\kappa_\textrm{gas}(T)$ and $\kappa_\textrm{holder}(T)$) and subjected to a heat flux boundary condition. 
As shown in~\cite{kappa_powders}, the effect thermal conductivity of the catalyst sample $\kappa^\textrm{eff}_\textrm{pellet}$ can be well described by the Maxwell-Garnett model~\cite{birdtransport,Pietrak-eff-kappa-compos-2014}, namely, 
\begin{align}\label{suppeq:kappa_cata}
\kappa^\textrm{eff}_\textrm{cata} = \kappa_\textrm{gas} + \dfrac{3\kappa_\textrm{gas}(f_\textrm{m} + f_\textrm{oxide})}{\dfrac{\kappa_\textrm{solid} + 2\kappa_\textrm{gas}}{\kappa_\textrm{solid} - \kappa_\textrm{gas}} - (f_\textrm{m} + f_\textrm{oxide})},
\end{align}
where $f_\textrm{m}$ and $f_\textrm{oxide}$ are the metal and the oxide volume fraction in the catalyst pellet, respectively. $\kappa_\textrm{gas}$ is the thermal conductivity of the gas and $\kappa_\textrm{solid}$ is the effective thermal conductivity of the solid material which itself can be again approximated by Maxwell-Garnett model~\cite{birdtransport,Pietrak-eff-kappa-compos-2014}
\begin{align}\label{suppeq:kappa_solid_mix}
\kappa_\textrm{solid} = \kappa_\textrm{oxide} + \dfrac{3\kappa_\textrm{oxide}\dfrac{f_m}{f_\textrm{m} + f_\textrm{oxide}}}{\dfrac{\kappa_\textrm{m} + 2\kappa_\textrm{oxide}}{\kappa_\textrm{m} - \kappa_\textrm{oxide}}-\dfrac{f_m}{f_\textrm{m} + f_\textrm{oxide}}},
\end{align}
where $\kappa_\textrm{m}$ and $\kappa_\textrm{oxide}$ are the metal thermal conductivity and the oxide thermal conductivity, respectively. The volume fraction of the metal and of the oxide can be deduced from the mass ($m_\textrm{cata}$) and the volume $V_\textrm{cata}$ of the pellet,
\begin{align}\label{suppeq:volume_fraction}
\begin{cases}
f_\textrm{m} = \dfrac{m_\textrm{cata}w_\textrm{m}/\rho_\textrm{m}}{V_\textrm{cata}},\\
f_\textrm{m}\rho_\textrm{m} + f_\textrm{oxide}\rho_\textrm{oxide} + \left(1 - f_\textrm{m} - f_\textrm{oxide}\right)\rho_\textrm{gas} = \dfrac{m_\textrm{cata}}{V_\textrm{cata}},
\end{cases}
\end{align}
where $w_m$ is the weight percentage of the metal, $\rho_\textrm{m}$, $\rho_\textrm{oxide}$ and $\rho_\textrm{gas}$ are the mass densities of the metal, of the oxide and of the gas, respectively. Since the metal volume fraction is usually much smaller than the oxide volume fraction {\XYZ (i.e. $f_\textrm{m} \ll f_\textrm{oxide}$)}, the effective thermal conductivity of the solid material is dominated by the thermal conductivity of the oxide {\XYZ (i.e. $\kappa_\textrm{solid} \approx \kappa_\textrm{oxide}$)}, justifying the ignorance of the metal NPs in Section~\ref{sec:methodology}. {\XYZ More importantly, since the metal nanoparticles and the metal oxide microparticles have a very small occupation fraction (i.e. $f_\textrm{m}+f_\textrm{oxide} \ll 1$) and since the thermal conductivity of the oxide is much larger than that of the gas (i.e. $\kappa_\textrm{oxide} \gg \kappa_\textrm{gas}$), Eq.~\eqref{suppeq:kappa_cata} becomes 
\begin{align}\label{suppeq:kappa_cata_approx}
\kappa^\textrm{eff}_\textrm{cata} \approx \left(1+3f_\textrm{oxide}\right)\kappa_\textrm{gas}
\end{align}
so that the thermal conductivity of the pellet is very close to that of the gas mixture and is much smaller than that of the oxide, as shown in Fig.~\ref{suppfig:kappa_cata_pellet}\footnote{The thermal conductivities of MgO and of Al\textsubscript{2}O\textsubscript{3} are taken from~\cite{kappa_MgO_2013} and~\cite{kappa_Al2O3_2014}, respectively.}.
}

The thermal conductivity of gases can be understand using the kinetic theory of gases~\cite{Francis-thermodynamics-kinetic-theory-book}. For ideal gases, the thermal conductivity is given by \begin{align}\label{suppeq:kappa_ideal_gas}
\kappa_\textrm{ideal gas} = \dfrac{f}{3D^2 }\sqrt{\dfrac{k_B^3 T}{\pi^3 M}},
\end{align}
where $f$ is the number of degrees of freedom, $D$ is the collision diameter, $M$ is the molecule mass and $k_B$ is the Boltzmann constant. Thus, the thermal conductivity of ideal gases increases with the temperature because the gas molecules collide with each other and transfer energy more frequently at high temperatures. For the same reason, the thermal conductivity of gases with lighter molecules is higher than that of gases with heavier molecules. Moreover, gases with larger molecules have lower thermal conductivity than gases with smaller molecules because of their larger collision diameter. When the gas is a mixture of reactants, products and carrier gas, the thermal conductivity of the gas mixture can be calculated using $\kappa_\textrm{gas mixture} = \displaystyle\sum_i x_i \kappa_i$, where $x_i$ and $\kappa_i$ are the mole fraction and the thermal conductivity of the $i$-th gas.

Next, we need to understand the efficiency of the sampler holder in releasing heat. To do that, we discuss separately the two experimental conditions - when the catalyst sample is not being {\XYZ resistive-heated} and when it is. 

For experiments where the catalyst sample is {\XYZ resistive-heated}, the sample holder temperature is controlled by the heater which is connected to the sample holder. In that sense, the sample holder serves only as a sort of ``boundary condition'' and its thermal conductivity is of no consequence. For the same reason, in this case the temperature dependence of the thermal conductivity of the sample holder cannot play any significant role on the photothermal nonlinearity.

\begin{figure}[h]
\centering
\includegraphics[width=0.7\textwidth]{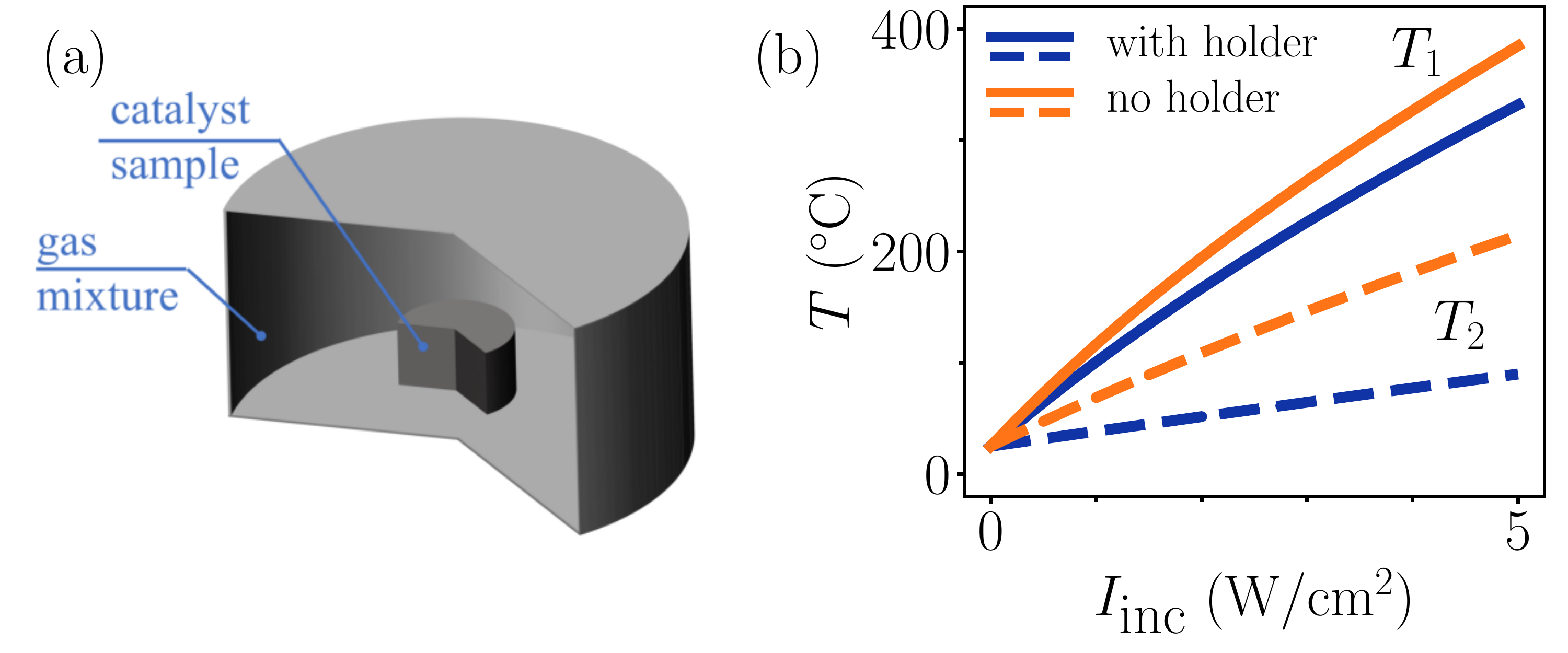}
\caption{(a) Schematic illustrations of the photocatalytic chamber without considering the steel holder, i.e. the photocatalyst sample is levitating. (b) $T_1$ and $T_2$ as a function of the illumination intensity. The blue solid line ($T_1$) and the blue dashed line ($T_2$) are the same as Fig.~3(b), representing the results of the simulation where the holder was included. The orange dotted solid line ($T_1$) and the dash-dotted line ($T_2$) representing the results of the simulation where the sample is levitating (no holder).}
\label{suppfig:T1_T2_supp_vs_levi}
\end{figure}

For experiments where the catalyst sample is not 
{\XYZ resistive-heated}, since the heat generation occurs primarily on the top layer of the catalyst, i.e., away from most of the sample holder, the sample holder plays a minor role. {\XYZ To see that, we performed a simulation where no holder is included and the sample is levitating as shown in Fig.~\ref{suppfig:T1_T2_supp_vs_levi}(a). The simulation results of $T_1$ and $T_2$ for the levitating sample are shown in Fig.~\ref{suppfig:T1_T2_supp_vs_levi}(b) and are compared with the results of the simulation where the holder was included. The simulation results show that when the holder is included, for the illumination intensity of 5 W/cm\textsuperscript{2} $T_2$ is reduced from $\sim 215 ^\circ$C to $\sim 90 ^\circ$C whereas $T_1$ is reduced from $\sim 385 ^\circ$C to $\sim 310 ^\circ$C. This indicates that the sample holder is effective to reduce the bottom-surface temperature but is less effective to reduce the top-surface temperature.} Moreover, since the increase of the holder temperature (which is almost equal to $T_2$) is much smaller than the that of the top-surface temperature $T_1$ when the illumination intensity increases (see Fig.~\ref{fig:JieLiu_NanoLett_2019_temp_map_T1_T2}(b)), and since the temperature dependence of the thermal conductivity of the sample holder~\footnote{The thermal conductivity of the steel support is~\cite{kappa_steel} $\kappa_{\textrm{steel}}(T)  = 14.6 + 1.27 \times 10^{-2} T$, where $\kappa_{\textrm{steel}}$ is in W/(m$\cdot$K) and $T$ is in $^\circ$C.} is 2.5 to 4 times weaker than that of gases, the sample holder plays only a minor role on the photothermal nonlinearity.


\begin{figure}[h]
\centering
\includegraphics[width=0.9\textwidth]{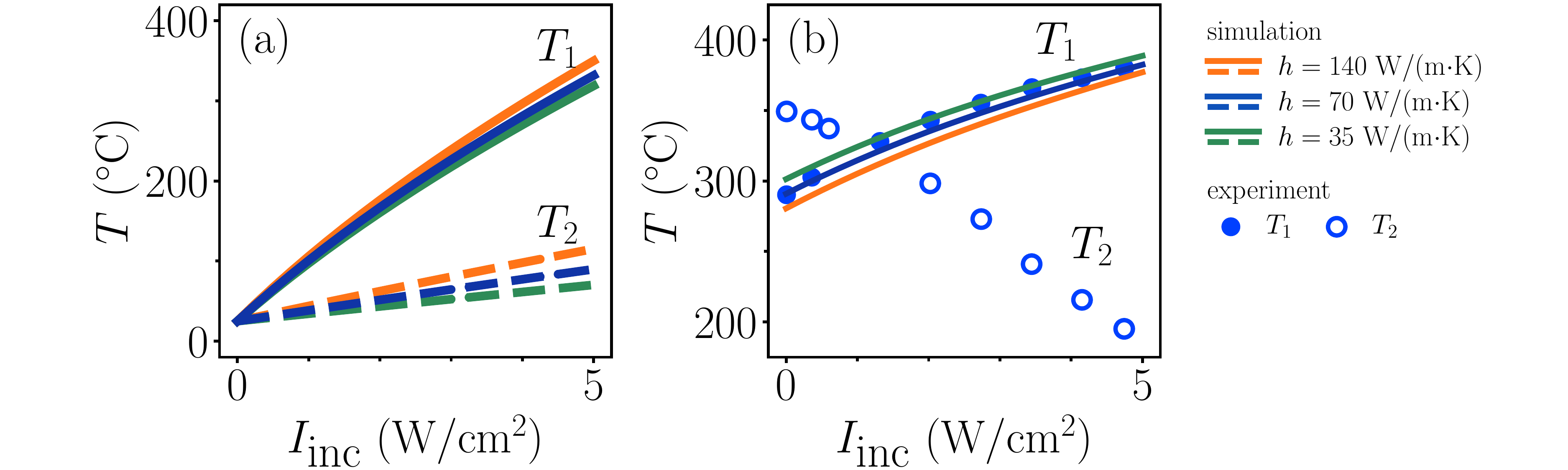}
\caption{(Color online) (a) $T_1$ and $T_2$ as a function of the illumination intensity for the case where the sample is not resistive-heated. The blue solid line ($T_1$) and the blue dashed line ($T_2$) are the same as Fig.~\ref{fig:JieLiu_NanoLett_2019_temp_map_T1_T2}(b), represent the results of the simulation where $h$ is set to be 70 W/(m$^2 \cdot$K). The orange (green) solid line and the orange dashed (green) line represent the results of the simulation where $h$ is set to be 140 W/(m$^2 \cdot$K) (35 W/(m$^2 \cdot$K)). (c) the same as (b), $T_1$ and $T_2$ as a function of the illumination intensity for the case where the sample is resistive-heated and is illuminated by the blue light source. }
\label{suppfig:JieLiu_NanoLett_2019_temp_map_T1_T2_h_para}
\end{figure}

Finally, the heat transfer from the chamber to the outer environment is modelled using a heat flux boundary condition driven by the temperature difference between the chamber boundary and the outer environment $h (T_{\textrm{boundary}} - T_{\textrm{env}})$. Therefore, a smaller value of $h$ gives rise to a stronger linear photothermal response. Based on this understanding, one can determine the value of $h$ by fitting the simulation results to the results of the experiments where the catalyst sample is not being 
{\XYZ resistive-heated} and small intensity limit, see e.g. Fig.~\ref{fig:JieLiu_NanoLett_2019_temp_map_T1_T2}(b). {\XYZ In Fig.~\ref{suppfig:JieLiu_NanoLett_2019_temp_map_T1_T2_h_para}, we plot the simulation results for the experiment~\cite{Liu-Everitt-Nano-Letters-2019} using different values of $h$ (35 W/(m$^2\cdot$ K), 70 W/(m$^2\cdot$ K) and 140 W/(m$^2\cdot$ K)). One can see that the simulation results of $T_1$ and $T_2$ are weakly sensitive to the value of $h$.} Moreover, the value of $h$ ensures that the rate of the illumination energy absorbed by the catalyst sample is equal to the rate of the thermal energy released to the outer environment through the boundary. For example, when the illumination intensity is 1 W/cm\textsuperscript{2}, the side boundary temperature is around 1$^\circ$C higher than the $T_{\textrm{env}}$ whereas the average temperature of the top- and of bottom-boundary are around 5.5$^\circ$C higher than the $T_{\textrm{env}}$, so that the releasing rate of the thermal energy is around $h A_{\textrm{side}} (T_{\textrm{side}} - T_{\textrm{env}}) + h A_{\textrm{top}}(T_{\textrm{top}} - T_{\textrm{env}}) + h A_{\textrm{bottom}} (T_{\textrm{top}} - T_{\textrm{env}}) \approx 0.285$W for $h \approx 70$ W/(m\textsuperscript{2}$\cdot$K), here $A_{\textrm{side}}$, $A_{\textrm{top}}$ and $A_{\textrm{bottom}}$ are the area of the side-, of the top- and of the bottom-boundary, respectively. This is in agreement with the rate of energy absorbed by the sample 0.283W. Moreover, once the value of $h$ is determined, it plays nearly no role on the nonlinear photothermal response. Finally, the value of $T_{\textrm{env}}$ was taken to be 20$^\circ$C. This is equal to the catalyst temperature when the catalyst sample is not {\XYZ resistive-heated} and there is no illumination.

\section{Supplemental simulation results for the experiments of Zhang \textit{et al.}~\cite{Liu_thermal_vs_nonthermal}}\label{suppsec:JLiu2018}
\begin{figure}[h]
\centering
\includegraphics[width=0.7\textwidth]{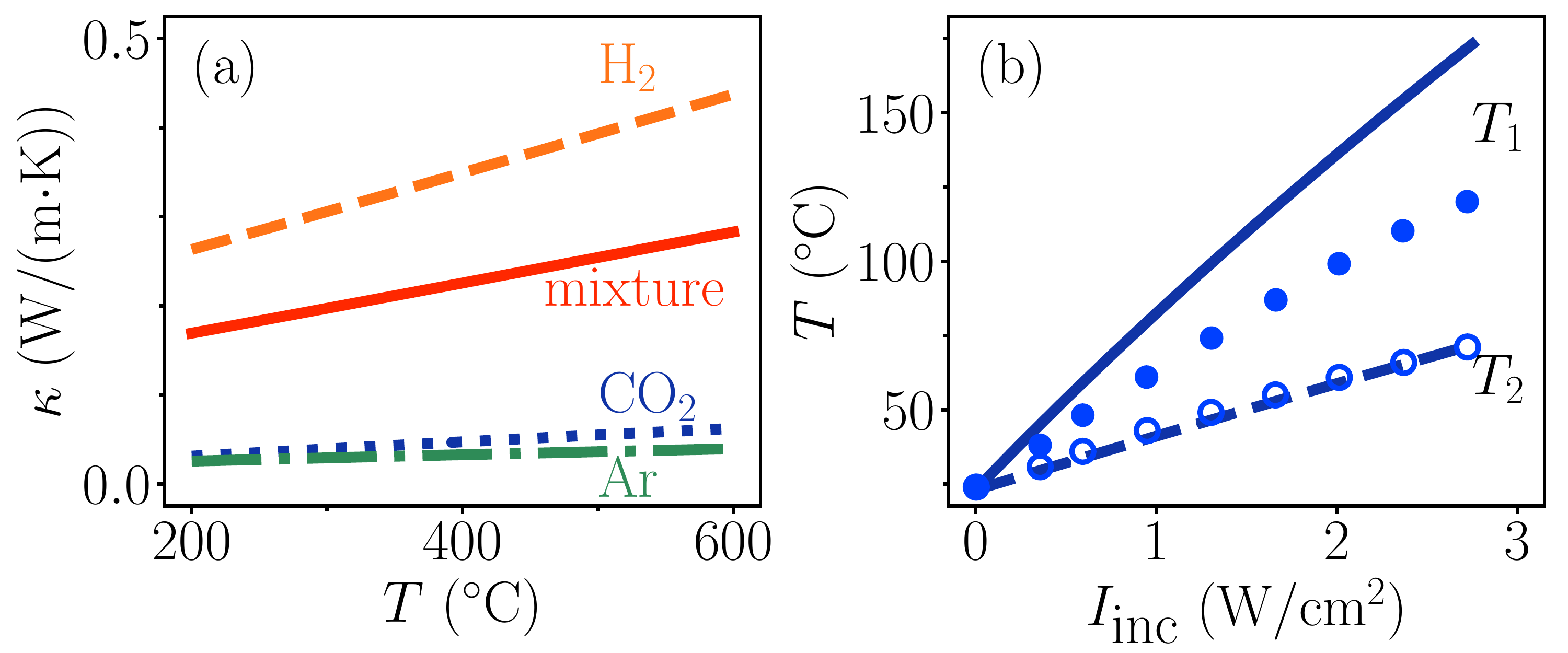}
\caption{(Color online) (a) Temperature dependence of the thermal conductivities of CO\textsubscript{2}~\cite{kappa_CO2_EngToolBox}, of H\textsubscript{2}~\cite{kappa_H2_EngToolBox}, Ar~\cite{kappa_Ar_Saxena_Chen_1975} and the mixture used in Ref.~\cite{Liu_thermal_vs_nonthermal}. (b) $T_1$ and $T_2$ as a function of the illumination intensity for the experiment without resistive heating in Ref.~\cite{Liu_thermal_vs_nonthermal}. The solid and dashed lines represent the COMSOL simulation results, and the symbols are extracted from Ref.~\cite{Liu_thermal_vs_nonthermal}.}
\label{suppfig:JieLiu_NanoLett_2018_T1_T2}
\end{figure}

{\XYZ We adapt the simulation configuration used in Section~\ref{sec:JLiu_nanolett} to the experimental setup described in Ref~\cite{Liu_thermal_vs_nonthermal} and simulate the temperature distribution using the heat transfer module of COMSOL Multiphysics. The thermal conductivity of the input gases~\cite{kappa_CO2_EngToolBox,kappa_H2_EngToolBox,kappa_Ar_Saxena_Chen_1975} and of their mixture are shown in Fig.~\ref{suppfig:JieLiu_NanoLett_2018_T1_T2}(a). The catalyst sample was 1 mm thick and was illuminated by a UV light source (365 nm). The penetration (skin) depth to the sample is calculated using the permittivity data of Rh~\cite{Adachi_book_optical_const_metal}. We again find that the penetration depth is much smaller than the sample thickness.}

{\XYZ Fig.~\ref{suppfig:JieLiu_NanoLett_2018_T1_T2} shows the simulation results of the top ($T_1$) and bottom ($T_2$) surface temperatures for the experiments without the resistive heating in Ref.~\cite{Liu_thermal_vs_nonthermal}. One can see that the simulation result of $T_1$ is around 50$^\circ$C higher than the experimental data at the illumination intensity of 2.75 W/cm\textsuperscript{2}. Since a large gas flow (250 sccm) was used in the experiment~\cite{Liu_thermal_vs_nonthermal} (which is $>3$ times larger than that (75 sccm) used in~\cite{Liu-Everitt-Nano-Letters-2019}), the gas mixture becomes more efficient in releasing heat. In this respect, using the non-convection approximation is one possible reason for the small between the simulation results and the experimental data. }

\section{Description of the COMSOL simulation file}
{\XYZ In this section, we provide the detailed description of the attached COMSOL simulation file. As mentioned in the main text, the simulation includes the catalyst sample, the sample holder and the reaction chamber with a few simplification. The geometry parameters are listed in the ``Description'' column of the ``Global definition - Parameters'' section. The temperature dependent thermal conductivity of the gas mixture (the red line in Fig.~\ref{fig:Halas_Science_2018_k_dsk_temp}(a)) is fitted using a polynomial of degree 3 and then is imported to the ``Component 1 - Materials'' section. The temperature dependent thermal conductivity of the catalyst sample is defined based on Eq.~\eqref{eq:kappa_cata_eff}. The illumination induced heat sources as a function of $z$ (normalized to 1 W/m\textsuperscript{2}) are calculated using Eq.~\eqref{eq:pabs}~\footnote{light source spectra are extracted from Fig.~S2 in Ref.~\cite{Liu-Everitt-Nano-Letters-2019}.} and are defined in the ``Global definition'' section as ``htsc365'', ``htsc455'', ``htscNIR'' and ``htscWHT'' for the UV (365 nm), blue (455 nm), NIR and white light sources, respectively. The heat source is imposed to the catalyst sample by setting the ``General source'' to be ``$\textrm{ity}*\textrm{htscXXX}(z)$'' in ``Component 1 - Heat transfer in Fluids - Heat source 1'' section, where ``$\textrm{ity}$'' is the illumination intensity and ``$\textrm{htscXXX}$'' is ``htsc365'', ``htsc455'', ``htscNIR'' or ``htscWHT''. In addition, a temperature boundary condition ``T2'' is set to control the bottom temperature of the catalyst sample to mimic the resistive heating, as mentioned in the main text. One needs to enable/disable the temperature boundary condition ``T2'' when running the simulation that the catalyst sample is/is not under resistive heating. ``Study Blue 455'', ``Study UV 365'', ``Study NIR'' and ``Study WHT'' are set in the model to simulate the temperature distribution for the blue, UV, NIR and white light sources, respectively. In each ``Study'', a ``Parametric Sweep'' is set to vary the illumination intensity ``$\textrm{ity}$''. When running the simulation for the catalyst sample under resistive heating, the ``Parametric Sweep'' is also set to vary ``$T_2$'' simultaneously. Here, $T_2$ is extracted from Fig.~S6 in Ref.~\cite{Liu-Everitt-Nano-Letters-2019}. When running the simulation for different light sources, one needs to use the corresponding heat source function ``$\textrm{htscXXX}$'' in ``Heat source 1'' section.} 

\providecommand*{\mcitethebibliography}{\thebibliography}
\csname @ifundefined\endcsname{endmcitethebibliography}
{\let\endmcitethebibliography\endthebibliography}{}

\bibliographystyle{rsc}